\documentclass[12pt]{article}
\usepackage[utf8]{inputenc}
\usepackage{amsmath,amssymb,amsfonts,amsthm}
\usepackage{graphicx,subfigure,color,psfrag}
\usepackage{bbm,cite}

\usepackage{psfrag}

\title{KPZ universality class and the anchored Toom interface}
\author{G.T. Barkema\thanks{Institute for Theoretical Physics, Universiteit Utrecht, Leuvenlaan 4, 3584 CE Utrecht, The Netherlands,
and Instituut-Lorentz, Universiteit Leiden, Niels Bohrweg 2, 2333 CA Leiden, The Netherlands. E-mail: {\tt G.T.Barkema@uu.nl}} \and P.L. Ferrari\thanks{Institute for Applied Mathematics, Bonn University, Endenicher Allee 60, 53115 Bonn, Germany. E-mail: {\tt ferrari@uni-bonn.de}} \and J.L. Lebowitz\thanks{Departments of Mathematics and Physics, Rutgers University, NJ 08854-8019, USA and Institute for Advanced Study, Einstein Drive, Princeton, NJ 08540, USA. E-mail: {\tt lebowitz@math.rutgers.edu}} \and H. Spohn\thanks{Zentrum Mathematik, TU M\"unchen, Boltzmannstrasse 3, D-85747 Garching, Germany and Institute for Advanced Study, Einstein Drive, Princeton NJ 08540, USA. E-mail: {\tt spohn@ma.tum.de}}}
\date{15. September 2014}

\begin{document}
\sloppy
\maketitle

\begin{abstract}
We revisit the anchored Toom interface and use KPZ scaling theory to argue that the interface fluctuations are governed by the Airy$_1$ process with the role of space and time interchanged. The predictions, which contain no free parameter, are numerically well confirmed for space-time statistics in the stationary state. In particular the spatial fluctuations of the interface computed numerically agree well with those given by the GOE edge distribution of Tracy and Widom.
\end{abstract}

\section{Introduction}\label{sec1}
Toom~\cite{T80} studied a family of probabilistic cellular automata on $\mathbb{Z}^2$ which have a unique stationary state at high noise level and (at least) two stationary states for low noise. Most remarkably, the low noise states are stable against small changes in the update rules~\cite{BG85}. This is in stark contrast to models satisfying the condition of detailed balance. For example the two-dimensional (2D) ferromagnetic Ising model with Glauber spin flip dynamics at sufficiently low temperatures and zero external magnetic field, $h=0$, has two equilibrium phases with non-zero spontaneous magnetization. But by a small change of $h$ uniqueness is regained~\cite{HS77}.

We consider the 2D Toom model with NEC (North East Center) majority rule. The system consists of Ising spins ($S_{i,j} = \pm 1$) located on a square lattice which evolve in discrete time. (We use magnetic language only for convenience. In physical realizations $S_{i,j}$ is a two-valued order parameter field). At each time step, all spins
$S_{i,j}$ are updated
independently according to the rule
\begin{equation}\label{1.1a}
S_{i,j}(t+1) =
 \begin{cases}
 \mathrm{sign}\big(S_{i,j+1}(t) + S_{i+1,j}(t) +S_{i,j}(t)\big) & \quad \mathrm{with\,\, probability\,\,} 1 - p - q\,,\\
 +1 & \quad \mathrm{with\,\, probability\,\,} p\,,\\
 -1 & \quad \mathrm{with\,\, probability\,\,} q\,.
 \end{cases}
\end{equation}
For $p=q=0$ we have a deterministic evolution: each updated spin becomes equal to the majority of itself and of its northern and eastern neighbors. Non-zero $p,q$ represents the effect of a noise which favors the $+$ sign
with probability $p$ and the $-$ sign with probability $q$. It was proved by Toom that for low enough noise ($p,q$ sufficiently small) the automaton has at least two translation invariant stationary states, such that the spins are predominantly $+$ or $-$, respectively. The probability with which one is obtained depends on the initial conditions.

\begin{figure}[h!]
\begin{center}
\psfrag{N}[cc]{$N$}
\psfrag{E}[cc]{$E$}
\psfrag{W}[cc]{$W$}
\psfrag{S}[cc]{$S$}
\psfrag{1}[cc]{$1$}
\psfrag{lambda}[cc]{$\lambda$}
\psfrag{n}[cc]{$n$}
\psfrag{h}[cc]{$\;\;M_n$}
\psfrag{-}[cc]{$-$}
\psfrag{+}[cc]{$+$}
\includegraphics[height=6cm]{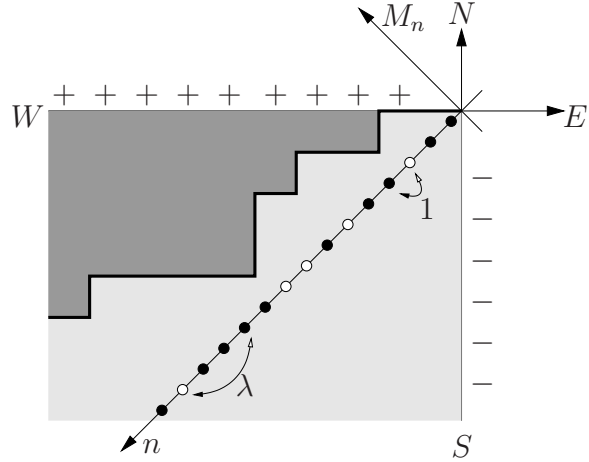}
\caption{Representation of the Toom interface model. The black/white dots are the spin values $+/-$ in the Toom spin exchange model.}
\label{Fig1}
\end{center}
\end{figure}

To investigate the spatial coexistence of the two phases, specific boundary conditions were introduced in~\cite{DLLS91,BBLS96}. More concretely, the Toom model restricted to the third quadrant was studied with the boundary conditions $S_{i,0} = 1$ and $S_{0,j} = -1$ for all $i,j<0$ and all $t$. Since the information is traveling southwest, in the long time limit a steady state is reached, for which the upper part is in one phase and the lower half in the other one. The phases
are bordered by an interface which fluctuates but has a definite slope, depending on $p,q$, on the macroscopic scale. Of interest are steady state static and dynamical fluctuations of this non-equilibrium interface.
Since both pure phases have already a nontrivial intrinsic structure, to analyse properties of the interface seems to be a difficult enterprise.
In~\cite{DLLS91,BBLS96} a low noise approximation is used for which the interface is governed by an autonomous stochastic dynamics in continuous time, see Figure~\ref{Fig1}. The interface can be represented by a spin configuration on the semi-infinite lattice $\mathbb{Z}_+$. Such spin configurations inherit then a dynamics in which spins are randomly exchanged. It is this Toom spin exchange model described below which is the focus of our contribution.
For more information we refer to~\cite{DLLS91,BBLS96}.

\subsection*{Toom spin exchange model}
We consider the 1D lattice $\mathbb{Z}$ and spin configurations $\{\sigma_j, j \in \mathbb{Z}, \sigma_j = \pm1\}$.
A \mbox{$+$ spin} exchanges with the closest \mbox{$-$ spin} to the right at rate $\lambda$ and, correspondingly, a \mbox{$-$ spin} exchanges with the closest $+$ spin to the right at rate $1$. $\lambda \in [0,1]$ is an asymmetry parameter.
The Bernoulli measures are stationary under this dynamics and we label them by their average magnetization, $\mu = \langle \sigma_0\rangle_\mu$.
On a finite ring of $N$ sites the dynamics is correspondingly defined, replacing right by clockwise. As can be seen from Figure~\ref{Fig1}, the interface is enforced by a hard wall at $0$, that is, spin configurations are restricted to the half lattice $\mathbb{Z}_+ = \{1,2,\ldots \}$, but the dynamics remains unaltered.
 The Toom spin model on the half lattice has an unusual independence property. If one considers the dynamics of the subsystem $\{\sigma_1(t),\ldots,\sigma_L(t)\}$, then it evolves as a continuous time Markov chain. However the magnetization is no longer conserved. If, for some $j$, the entire block $[j,\ldots,L]$ has spin $+$,
 then $\sigma_j(t)$ flips to $- \sigma_j(t)$ with rate $\lambda$ and correspondingly for a block of \mbox{$-$ spins} touching the right border the flip is done with rate 1. As a consequence, a unique limiting probability measure is approached as $t \to \infty$.
 In our approximation, the height of anchored interface of the Toom automaton is just the magnetization of the Toom spin model,
\begin{equation}\label{1.1}
M_n(t) = \sum_{j=1}^n \sigma_j(t)\,.
\end{equation}
The argument $t$ is omitted in case the $n$-dependence at fixed $t$ is considered. Averages in the steady state are denoted by $\langle \cdot \rangle$. Note that by time stationarity \mbox{$\langle M_n(t) \rangle = \langle M_n(0) \rangle = \langle M_n \rangle$} and time correlations such as $\langle M_n(t) M_{n'}(t')\rangle$ depend only on $t - t'$. At $\lambda = 1$ the interface is along the diagonal and fluctuates symmetrically, $\langle M_n \rangle = 0$, while for $0 < \lambda < 1$ the interface becomes asymmetric.

Based on theoretical and numerical evidence, in~\cite{DLLS91} it was concluded that, for large $n$,
\begin{equation}\label{1.2}
 \langle M_n^2 \rangle - \langle M_n \rangle^ 2\simeq n^{1/2} \quad\mathrm{for} \,\,\lambda = 1
\end{equation}
with possibly logarithmic corrections, while
\begin{equation}\label{1.3}
 \langle M_n^2 \rangle - \langle M_n \rangle^ 2\simeq n^{2/3} \quad\mathrm{for} \,\,0 < \lambda < 1 \,.
\end{equation}
Most remarkably, using the then just being developed multi-spin coding techniques, the full probability density function (pdf) for $M_n$ was recorded, see~\cite{BBLS96}, Fig. 3. For $\lambda = 1$, the pdf is well fitted by a Gaussian,
in agreement with the prediction of the collective variable approximation (CVA)~\cite{DLLS91}. The variance differed however by a logarithmic correction from the $\sqrt{n}$ prediction, in the scaling limit, given by the CVA. For $\lambda = \tfrac{1}{4}$, the scaling function obtained through the CVA was used as a fit to the numerical data. This
is given by $\mathrm{Ai}(x)^4$ with $\mathrm{Ai}$ the standard Airy function. Somewhat \textit{ad hoc}, the left tail of
$\mathrm{Ai}$ was cut at its first zero. Looking eighteen years later at the same figure, with the hindsight of the much improved
understanding of the KPZ universality class, it is a safe guess that in fact a Tracy-Widom distribution from random matrix theory
is displayed. Apparently the fluctuations of the anchored Toom interface share the same fate as the length of the longest increasing subsequence of random permutations. Without knowing, Odlyzko~\cite{OR99} observed the GUE Tracy-Widom distribution.
We refer to~\cite{HL14} for a more complete account of the history. For us Fig. 3 of~\cite{BBLS96} is a compelling motivation to return to the fluctuations of the anchored Toom interface and to understand better how they fit into the KPZ universality class.

In this note, we will provide numerical and theoretical evidence that in fact
\begin{equation}\label{1.4}
M_n \simeq \mu_0 n + (\Gamma n )^{1/3}\tfrac{1}{2} \xi_{\mathrm{GOE}}
\end{equation}
for large $n$ and $0 < \lambda < 1$. Here the coefficients $\mu_0,\Gamma$ depend on $\lambda$ and are computed explicitly. The random amplitude $\xi_{\mathrm{GOE}}$ is GOE Tracy-Widom distributed. The general form of (\ref{1.4}) is familiar from other models in the KPZ universality class. To have fluctuations governed by the GOE edge distribution came as a complete surprise and has not been anticipated before. To be on the safe side we also investigate the covariance
$\langle M_n(t) M_{n}(0)\rangle - \langle M_n(0)\rangle^2$ and compare it with the prediction coming from the covariance of the Airy$_{1}$
process. Besides running multi-spin coding on more modern machines, we present a much improved
analysis on interchanging the role of space and time for the interface dynamics.

%%%%%%%%%%%%%%%%%%%%%%%%%%%%%%%%%%%%%%%%%%%%%%%%%%%%%%%%%%

\section{Mesoscopic description of the Toom interface}\label{sec2}
To study the fluctuations of the Toom interface, it is convenient to start from a mesoscopic description of the height
\begin{equation}\label{2.3}
h(x,t) \simeq M_n(t)\,,
 \end{equation}
where $x$ stands for the continuum approximation of $n$. Firstly note that on $\mathbb{Z}$ the Toom spin model conserves the magnetization and thus has a one-parameter family of stationary states labeled by the average magnetization,
$\mu$. In the steady state the spins are independent and the spin current is given by
\begin{equation}\label{2.1}
J(\mu,\lambda) = 2 \Big( \lambda \frac{1+\mu}{1-\mu} - \frac{1-\mu}{1+\mu}\Big)\,,
\end{equation}
see~\cite{DLLS91}. For the anchored Toom interface we expect (and have checked numerically) that in small segments very far away from the origin the spins will be independent, so that, $\langle \sigma _i\sigma_{i+j} \rangle - \langle \sigma _i\rangle \langle\sigma_{i+j}\rangle \to 0$ as $i \to \infty$ at fixed $j \neq 0$.
To have a stationary state for the semi-infinite system thus requires $J =0$. Using (\ref{2.1}) this has the unique solution
\begin{equation}\label{2.2}
\mu_0 = \frac{1-\sqrt{\lambda}}{1 +\sqrt{\lambda}}\,,
\end{equation}
which determines the asymptotic magnetization. If $h$ is slowly varying on the scale of the lattice, then locally it will maintain a definite slope $u = \partial_x h$. The local slope is conserved, hence governed by the conservation law
\begin{equation}\label{2.3a}
\partial_t u + \partial_x J(u,\lambda) = 0\,,
 \end{equation}
which should be viewed as the Euler equation for the magnetization of the spin model. Equivalently, there is a Hamilton-Jacobi type equation for $h$,
\begin{equation}\label{2.3c}
\partial_t h + J(\partial_x h,\lambda) = 0\,,
 \end{equation}
with $h(x) = \mu_0x$ as stationary solution.

To describe on a mesoscopic scale the statistical properties of the Toom interface, we follow the common practice to add noise to the deterministic equation (\ref{2.3c}), see for example~\cite{LL}. More concretely to the spin current in (\ref{2.3a}) we add the dissipative term $-\tfrac{1}{2}D\partial_x u$, $D$ the diffusion constant, and, since local exchanges are essentially uncorrelated, the space-time white noise $\kappa W(x,t)$.
$W(x,t)$ is normalized and $\kappa$ is the noise strength. Since the deviation from the constant slope profile $\mu_0 n$
will be studied, in fact it suffices, by power counting, to keep the current $J(\mu,\lambda)$ up to second order relative to
$\mu_0$ as
\begin{equation}\label{2.3b}
J(\mu -\mu_0, \lambda ) = v(\lambda) (\mu -\mu_0)+\tfrac{1}{2}G(\lambda)(\mu -\mu_0)^2 + \mathcal{O}((\mu -\mu_0)^3)\,,
\end{equation}
where
\begin{equation}\label{2.5}
v(\lambda) = 2(1+ \sqrt{\lambda})^2\,,\quad G(\lambda) = (1+ \sqrt{\lambda})^3(1 - \sqrt{\lambda})\frac{1}{\sqrt{\lambda}}\,.
\end{equation}
Thereby one obtains that on a mesoscopic scale the fluctuating height $h(x,t)$ is governed by
 \begin{equation}\label{2.6}
\partial_t h = - v\partial_x h - \tfrac{1}{2}G(\partial_x h)^2 + \tfrac{1}{2}D\partial_x^2h + \kappa W(x,t)
\end{equation}
for $t \geq 0$. For the Toom interface the height is pinned at the origin which leads to the restriction $x \geq 0$ and the boundary condition
\begin{equation}\label{2.2c}
h(0,t) = 0\,.
\end{equation}
The coefficients $v,G$ depend on $\lambda$. If $G =0$, as is the case for $\lambda = 1$, the second order expansion
does not suffice. Fourth order is irrelevant, but the third order term generates logarithmic corrections~\cite{P},
which are the theoretical reason behind the already mentioned logarithmic corrections for the interface variance at $\lambda = 1$.

Eq.~(\ref{2.6}) is the much studied one-dimensional KPZ equation~\cite{KPZ} with two important differences. Firstly the height function is over the half-line, being pinned at the origin, and secondly there is the outward drift $v(\lambda)$, which cannot be removed because of this boundary condition.
At such brevity our reasoning may look \textit{ad hoc}. But the scheme should be viewed as a particular case of the KPZ scaling theory~\cite{KHM,S}, which has been confirmed through extensive Monte Carlo simulations for related models, for example see~\cite{BS}.

For magnetization $\mu$ the spin susceptibility $A$ equals, for independent spins, $\langle \sigma_0^2\rangle_\mu - \langle \sigma_0\rangle_\mu^2 = 1 -\mu^2$ and at $\mu_0$ is given by,
\begin{equation}\label{2.7}
A = 4 \sqrt{\lambda} (1+ \sqrt{\lambda})^{-2}\,.
\end{equation}
To connect with the parameters of Eq.~(\ref{2.6}), one checks that on $\mathbb{R}$ the steady state has the slope statistics given by spatial white noise with
variance $\kappa^2/D$. Therefore we identify as
\begin{equation}\label{2.6a}
A = \kappa^2/D\,.
\end{equation}
In the scaling regime only $A$ will appear, which is unambiguously defined by (\ref{2.7}) in terms of the spin model, while $D$ and $\kappa$ separately are regarded as phenomenological coefficients. As for $M_n(t)$, our focus is the stationary process
determined by (\ref{2.6}), (\ref{2.2c}).

%%%%%%%%%%%%%%%%%%%%%%%%%%%%%%%%%%%%%%%%%%%%%%%%%%%%%%%%
\section{Interchanging the role of space and time}\label{sec2a}
Considering Eq. (\ref{2.6}), it would be of advantage to interchange $x$ and $t$, because then the boundary value
$h(0,t) = 0$ turns into an initial condition, which is more accessible. From the perspective of stochastic partial differential equation, such an interchange
looks impossible. But once we write the Cole-Hopf solution of (\ref{2.6}), for example see~\cite{CH}, our scenario becomes fairly plausible.

The Cole-Hopf transformation is defined by
\begin{equation}\label{2.8}
Z(x,t) = \mathrm{e}^{(G/D)h(x,t)}\,,
\end{equation}
which satisfies
\begin{equation}\label{2.9}
\partial_tZ = \tfrac{1}{2} D\partial_x^2 Z -v\partial_x Z - (G\kappa/D) W Z
\end{equation}
on $\mathbb{R}_+$ with boundary condition $Z(0,t) = 1$ and some initial condition $Z_0(x)$.
The first two terms generate a Brownian motion with constant drift, which is used in the Feynman-Kac discretization to formally integrate (\ref{2.9}).
Let $b(t)$ be a Brownian motion with $b(0) = 0$ and variance variance $ \mathbb{E}(b(t)^2) = Dt$, $ D > 0$, $ \mathbb{E}$ denoting the expectation for $b(t)$. In the usual parlance $b(t)$ is called a directed polymer, since it moves forward in the time direction. Furthermore let $T$ be the largest $s$ such that $x + b(t-s) -v(t - s) =0$, i.e. $T$ is the first time of hitting of $0$ for a Brownian motion with drift starting at $x$. Then Eq.~(\ref{2.9}) integrates to
\begin{equation}\label{2.10}
Z(x,t) = \mathbb{E} \Big( \mathrm{e}^{-(G\kappa/D) \int_{(t - T) \vee 0}^t ds W(x + b(t-s) -v(t-s),s)}\big(Z_0(b(t) -vt)\mathbbm{1}_{[t \leq T]}
+ \mathbbm{1}_{[t > T]}\big)\Big)\,.
\end{equation}
For large $t$, the path $\{x + b(t-s) - v(t-s), 0 \leq s \leq t\}$ will hit 0 before $s = 0$ with a probability close to one. Hence the contribution from the term with $Z_0$ will vanish, the particular initial conditions are forgotten, and
\begin{equation}\label{2.11}
\lim_{t \to \infty} Z(x,t) = Z_{\infty}(x) = \mathbb{E}\Big( \mathrm{e} ^{-(G\kappa/D)\int^0_{-T} ds W(x +b(-s) +vs,s)}
\Big)\,.
\end{equation}
Going back to (\ref{2.8}), $(D/G)\log Z_{\infty}(x)$ defines the stationary measure for Eqs. (\ref{2.6}), (\ref{2.2c}). The stationary process for all $t \in \mathbb{R}$ is obtained by shifting $W$ in $t$ as
\begin{equation}\label{2.12}
Z_{\mathrm{st}}(x,t) = \mathbb{E}\Big( \mathrm{e} ^{-(G\kappa/D)\int^0_{-T} ds W(x + b(-s) +vs,s+t)}
\Big)\,.
\end{equation}

To understand the interchange between $x$ and $t$, at least in principle, we discretize (\ref{2.12}) by replacing $\mathbb{R}_+ \times
\mathbb{R}$ by $\mathbb{Z}_+\times\mathbb{Z}$. Then the continuum directed polymer $b(t) -vt$ is replaced by its discrete cousin, namely a random walk $\omega$
with down-left paths only. The walk starts at $\vec{j}_0$, $\vec{j} =(j_1,j_2)$. The transitions are $\omega_n$ to
$\omega_n - (1,0)$ with probability $p$ and $\omega_n$ to $\omega_n - (0,1)$ with probability $q$, $p+q = 1$.
$T$ is the time of first hitting the line $\{j_2 = 1\}$. $W(x,t)$ is replaced by a collection of independent standard Gaussian random variables $\{W(j_1,j_2),j_1\in\mathbb{Z}, j_2 \in \mathbb{Z}_+\}$. The integral in the exponent of (\ref{2.12}) now turns into the sum over $W(j_1,j_2)$ along $\omega$ until the boundary is reached. Since the path $\omega$ is decreasing, it can be viewed with either $j_1$ or $j_2$ as time axis. In the first version, the continuum limit equals $-(q/p)t + \sqrt{q}\,b(t)$ and in the second version $-(p/q)t + \sqrt{p}\,b(t)$.

Eq.~(\ref{2.12}) corresponds to the first version. Instead we now take $j_2$ as time axis and consider the continuum version of the partition function as in Eq.~(\ref{2.12}). Then the directed polymer is parametrized as
$u \mapsto t + \tilde{b}(x-u) - \tilde{v}(x - u)$. $\tilde{b}(u)$ is a Brownian motion with $\tilde{b}(0) = 0$ and variance $\tilde{\mathbb{E}}(\tilde{b}(u)^2) = \tilde{D} u $ and the transformed drift is $\tilde{v} = v^{-1}$. With these conventions the partition function reads \begin{equation}\label{2.12a}
\tilde{Z}_{\mathrm{st}}(x,t) = \tilde{\mathbb{E}} \Big( \mathrm{e} ^{(\tilde{G}\tilde{\kappa}/\tilde{D})\int_0^{x} du W(u, t + \tilde{b}(x-u) -\tilde{v}(x-u))}
\Big)\,,
\end{equation}
where $x>0$ and $t\in\mathbb{R}$. By defining $\tilde{h} = (\tilde{D}/\tilde{G})\log \tilde{Z}_{\mathrm{st}}$, one arrives at
\begin{equation}\label{2.20a}
\partial_x \tilde{h} = - \tilde{v}\partial_t \tilde{h} - \tfrac{1}{2} \tilde{G} (\partial_t \tilde{h})^2
+ \tfrac{1}{2}\tilde{D} \partial_t^2\tilde h + \tilde{\kappa} W
\end{equation}
with initial condition
\begin{equation}\label{2.13a}
\tilde{h}(0,t) =0\,.
\end{equation}

There is no good reason for having a strict identity between $h$ and $\tilde{h}$. But one would expect both to have the same asymptotic behavior, \textit{provided} one appropriately adjusts $\tilde{G}$ and $\tilde{A} = \tilde {\kappa}^2/\tilde{D}$. The argument given is not specific enough for finding out the correctly transformed coefficients. For this purpose
we return to the Toom spin model on $\mathbb{Z}$
and first consider the macroscopic height evolution. Then, as in (\ref{2.3a}),
\begin{equation}\label{2.13}
\partial_t h +J(\partial_x h) = 0\,,
\end{equation}
where $J(\partial_x h) = J(\partial_x h, \lambda)$. Since $J$ is monotone, it is invertible and
 \begin{equation}\label{2.14}
\partial_x h + \tilde{J}(\partial_t h) = 0\,,\quad J(\tilde{J}(u)) = u\,,
\end{equation}
and, expanding in $\partial_t h$,
\begin{equation}\label{2.15}
\partial_x h = - v^{-1}\partial_t h + \tfrac{1}{2} G v^{-3}(\partial_t h)^2 + \mathcal{O}((\partial_t h )^3)\,.
\end{equation}
We conclude that
 \begin{equation}\label{2.16h}
 v\tilde{v} = 1\,, \quad G = -\tilde{G} v^3\,.
\end{equation}

As a second task we have to find out the transformed susceptibility $\tilde{A}$. For this purpose we consider the stationary Toom spin model, $\sigma_j(t)$,
 on $\mathbb{Z}\times \mathbb{R}$ with average magnetization $\mu$. Since the steady state is Bernoulli,
 one already knows that
 \begin{equation}\label{2.17}
 \sum_{j\in\mathbb{Z}} \big(\langle \sigma_j(0)\sigma_0(0)\rangle _\mu - \mu^2\big) = 1 - \mu^2 = A\,.
 \end{equation}
$\tilde{A}$ is the corresponding susceptibility in the $t$-direction, which is defined by
\begin{equation}\label{2.16a}
\int_{-\infty}^{\infty} dt \big(\langle \sigma_0(t)\sigma_0(0)\rangle _\mu - \mu^2\big) = \tilde{A} \,.
\end{equation}
The computation of $\tilde{A}$ requires dynamical correlations, which looks like a difficult task. Help comes from the very special correlation structure which holds for
a large class of 1D spin models with exchange dynamics. While for some models such structure can be checked
from the exact solution~\cite{PrSp04,FeSp06}, for the Toom spin model it is an assumption. But there is no good
reason why the Toom spin model should behave exceptionally. We consider correlations between $(0,0)$ and
$(j,t)$. There then is a special direction, determined through the speed of propagation of small disturbances,
$v(\lambda)$. Along $(v(\lambda)s,s)$ the line integral as in (\ref{2.16a}) vanishes, while in all other directions
it converges to a strictly positive value.

For the complete argument it is convenient to first
define the height function for the Toom spin model by
\begin{equation}\label{2.18}
\mathsf{h}(j,t) = \begin{cases}
 \sum_ {i = 1}^j (\sigma_i(t) - \mu)
 & \quad \mathrm{for\,\, } j> 0\,,\\
 \mathcal{J}_{(0,1)}([0,t] ) - Jt& \quad \mathrm{for\,\, } j=0\,,\\
 \sum_ {i = j}^{-1} (\sigma_i(t) -\mu) & \quad \mathrm{for\,\, } j< 0\,.
 \end{cases}
\end{equation}
Here $ \mathcal{J}_{(0,1)}([0,t] )$ is the actual time-integrated spin current across the bond $(0,1)$ up to time $t$
implying the convention $\mathsf{h}(0,0) = 0$. By definition the spin susceptibility along the $j$-axis is given by
 \begin{equation}\label{2.19}
 \langle (\mathit{\mathsf{h}}(j,0) - \mathsf{h}(0,0) )^2\rangle = A j
\end{equation}
for large $j$, $j > 0$.
Correspondingly in the $t$-direction
 \begin{equation}\label{2.20}
 \langle (\mathsf{h}(0,t) - \mathsf{h}(0,0) )^2\rangle = \tilde{A} t
\end{equation}
for large $t$, $t > 0$. In the direction of the propagation speed, the height fluctuations of
are suppressed,
\begin{equation}\label{2.21}
 \langle (\mathsf{h}(vt,t) - \mathsf{h}(0,0) )^2\rangle = \mathcal{O}(t^{2/3})\,.
\end{equation}
We set $X= \mathsf{h}(vt,t) - \mathsf{h}(0,t)$ and $Y=\mathsf{h}(0,t)- \mathsf{h}(0,0)$. Using the general bound
$ |\langle X^2 \rangle - \langle Y^2 \rangle| \leq
\langle (X+Y)^2\rangle^{1/2} (2 \langle X^2 \rangle + 2 \langle Y^2 \rangle )^{1/2}$ and stationarity,
\begin{equation}\label{2.22}
\mathsf{h}(vt,t) - \mathsf{h}(0,t) = \mathsf{h}(vt,0) - \mathsf{h}(0,0)
\end{equation}
in distribution, one concludes that
\begin{equation}\label{2.23}
\lim_{t\to \infty} t^{-1} \langle (\mathsf{h}(vt,0) - \mathsf{h}(0,0) )^2\rangle
= Av = \tilde{A} = \lim_{t\to \infty} t^{-1} \langle (\mathsf{h}(0,t) - \mathsf{h}(0,0) )^2\rangle\,.
\end{equation}

Our argument only used that along a particular direction the height fluctuations are subdiffusive. While such a property
is expected to hold for a large class of spin exchange dynamics, it has been proved only for a few models, in particular
for the asymmetric simple exclusion process (ASEP)~\cite{LY04,BS10}. Here
particles hop to the right with rate $p$ and to the left with rate $q$, $p+q = 1$, provided the target site is empty.
As for the Toom spin model the invariant measures are Bernoulli, say with density $\rho$. Then $A = \rho (1 - \rho)$
and $v = (p-q)(1 -2\rho)$. The identity (\ref{2.23}) states that $\langle \big(\mathcal{J}_{(0,1)}([0,t] ) - \rho(1- \rho)t\big)^2\rangle = \tilde{A} t $ for large $t$ with $\tilde{A} = |(p-q)(1 - 2\rho)|\rho (1 - \rho)$. In fact, this identity is proved in
\cite{FF94}, including the corresponding central limit theorem.

%%%%%%%%%%%%%%%%%%%%%%%%%%%%%%%%%%%%%%%%%%%%%%%%%%%%%%%%
\section{Asymptotic properties}\label{sec3}
As argued in the previous section, on a large space time scale the stationary process $M_n(t) - \mu_0 n$ is
approximated by $\tilde{h}(x,t)$ governed by Eq.~(\ref{2.20a}) with initial conditions $\tilde{h}(0,t) = 0$,
which is known as KPZ equation with flat initial conditions. Available are a replica solution~\cite{CL13} and proofs for a few discrete models in the KPZ universality class~\cite{F04,Sas07,BFPS07,BFP07,BFP08,FSW14}. We summarize the findings, which then immediately yields the predictions for the anchored Toom interface.
The non-universal parameters are \mbox{$\tilde{v} = 2^{-1}(1+ \sqrt{\lambda})^{-2}$}, $\tilde{A} = 8 \sqrt{\lambda}$,
and $\tilde{G} = -2^{-3}(1+ \sqrt{\lambda})^{-3}(1 - \sqrt{\lambda})\frac{1}{\sqrt{\lambda}}$. Following~\cite{TS12} we introduce
\begin{equation}\label{3.1}
\tilde{\Gamma} = |\tilde{G}|\tilde{A}^2 = 8\sqrt{\lambda}(1 - \sqrt{\lambda})(1+ \sqrt{\lambda})^{-3}\,.
\end{equation}
Then, for large $x$,
\begin{equation}\label{3.2}
\tilde{h}(x,0) \simeq \tilde{v} x +
(\tilde{\Gamma} x)^{1/3} \tfrac{1}{2} \xi_{\mathrm{GOE}}\,,
\end{equation}
where the random amplitude $\xi_{\mathrm{GOE}}$ is GOE Tracy-Widom distributed. More precisely $\xi_{\mathrm{GOE}}$ has the distribution function
\begin{equation}\label{3.3}
\mathbb{P}(\xi_{\mathrm{GOE}} \leq s) = F_1(s)\,,\quad F_1(2s) = \det(\mathbbm{1} - K)_{L^2((s,\infty))}\,.
\end{equation}
The integral kernel of $K$ reads $K(u,u') = \mathrm{Ai}(u+u')$, see~\cite{FS06} for this particular representation of $F_1$.
 As a consequence, for large $n$, $M_n - \mu_0n $ is predicted to have the
distribution function
\begin{equation}\label{3.4}
\mathbb{P}(M_n - \mu_0 n \leq s) \simeq F_1(2(\tilde{\Gamma} n)^{-1/3}s)\,.
\end{equation}

$\frac12 \xi_{\rm GOE}$ has mean $-0.6033$, variance $0.408$, and decays rapidly at infinity as \mbox{$\exp[-2(2s)^{3/2}/3]$} for
the right tail and $\exp[-|s|^3/6]$ for the left tail. The GOE Tracy-Widom distribution was originally derived in the context of random matrices~\cite{TW94}. One considers the Gaussian orthogonal ensemble of real symmetric $N\times N$ matrices, $H$, with probability density
\begin{equation}\label{3.5}
Z^{-1} \exp\left(- \tfrac{1}{4N}\mathrm{tr}H^2\right) dH,
\end{equation}
where $dH=\prod_{1\leq i\leq j\leq N}dH_{i,j}$. Let $\lambda_N$ be the largest eigenvalue of $H$. Then, for large $N$,
\begin{equation}\label{3.6}
\lambda_N \simeq 2N + N^{1/3} \xi_{\mathrm{GOE}} \,.
\end{equation}

Next we consider $t \mapsto \tilde{h}(x,t)$ as a stationary stochastic process in $t$. It is correlated over times of order
$(\tilde{\Gamma} x)^{2/3}$. In fact after an appropriate scaling $\tilde{h}(x,t)$ converges to a stochastic process known as Airy$_1$. In formulas
\begin{equation}\label{3.7}
\lim_{x \to\infty} (\tilde{\Gamma} x )^{-1/3}\big(\tilde{h}(x, 2\tilde{A}^{-1} (\tilde{\Gamma} x )^{2/3}t) -\tilde{v}x\big)= \mathcal{A}_1(t)\,.
\end{equation}
For the joint distribution of $\mathcal{A}_1(t_1),\ldots,\mathcal{A}_1(t_n)$, $t_1 <\ldots< t_n$, one has a determinantal formula. In particular for two times $t_1,t_2$
\begin{equation}\label{3.8}
\mathbb{P}( \mathcal{A}_1(t_1) \leq s_1, \mathcal{A}_1(t_2) \leq s_2) = \det (\mathbbm{1} - \mathsf{K})_{L^2(\mathbb{R}\times\{1,2\})}.
\end{equation}
$\mathsf{K}$ is a operator with kernel given by
\begin{equation}
\mathsf{K}(x,i;x',j)=\mathbbm{1}(x > s_i) K_1(t_i,x;t_j,x')\mathbbm{1}(x' > s_j),
\end{equation}
with
\begin{equation}\label{3.8a}
\begin{aligned}
K_1(t,x;t',x') =& \mathrm{Ai}(x'+x+(t'-t)^2)\exp \big((t'-t)(s'+s)+\tfrac{2}{3}(t' - t)^3\big)\\
&-\,\frac{1}{\sqrt{4 \pi (t'-t)}}\exp\left( -\frac{(x'-x)^2}{4(t'-t)}\right) \mathbbm{1}(t'>t)\,.
\end{aligned}
\end{equation}
From the expression (\ref{3.8}) one obtains the covariance
 \begin{equation}\label{3.9}
 g_1(t) = \langle \mathcal{A}_1(0) \mathcal{A}_1(t)\rangle
 - \langle \mathcal{A}_1(0)\rangle^2\,.
 \end{equation}
 To actually compute $g_1$, one uses a matrix approximation of the operators in (\ref{3.8}) by evaluating the kernels
 at judiciously chosen base points~\cite{Bo08}, for which the determinants are then readily obtained by a standard numerical routine. The limit in (\ref{3.7}) implies that, for large $x$,
 \begin{equation}\label{3.10}
 \langle \tilde{h}(x,0) \tilde{h}(x,t)\rangle
 - \langle \tilde{h}(x,0)\rangle^2 \simeq (\Gamma x)^{2/3}g_1(\tilde{A}t/2(\tilde{\Gamma} x)^{2/3})\,.
 \end{equation}
 Returning to the Toom interface one arrives at the result that, for large $n$,
\begin{equation}\label{3.11}
\langle (M_n(t) - \mu_0 n)(M_n(0) - \mu_0 n) \rangle
 - \langle( M_n(0) - \mu_0 n)\rangle^2 \simeq (\tilde{\Gamma} n)^{2/3} g_1(\tilde{A}t/2(\tilde{\Gamma} n)^{2/3})\,.
\end{equation}
Based on KPZ scaling theory, (\ref{3.4}) and (\ref{3.11}) are our predictions for the fluctuations of the Toom interface. They will be tested numerically in the following section.

\newpage
%%%%%%%%%%%%%%%%%%%%%%%%%%%%%%%%%%%%%%%%%%%%%%%%%%%%%%%

\section{Numerical studies}\label{sec4}
The Toom spin model lends itself well for an efficient simulation technique,
often referred to as multispin coding~\cite{Newman99}, which was used already in~\cite{BBLS96} and is used also in this study. The basic idea
is that the time-consuming part of the algorithm is written down as a
sequence of single-bit operations, but the computer then acts on 64-bit
words, thereby performing 64 simulations simultaneously. Most of the
computational effort is invested into selecting a random site, flipping
the spin value at that site, and then walking along the array of spins
until an opposite spin is encountered, which is then also flipped. A
piece of code in the programming language C which achieves this is:

\begin{verbatim}
i=random()*n;
first=spin[i];
todo=randword()|first;
spin[i]^=todo;
for (j=i+1;(j<n)&&(todo!=0);j++)
{
 flip=todo&(first^spin[j]);
 spin[j]^=flip;
 todo&= (~flip);
}
\end{verbatim}
In this example code, the introduction of a random pattern {\bf randword()} in the third line introduces a bias; the density of 1s in this random pattern should equal~$\lambda$.

For the actual simulations, we start from a random spin distribution, that is, the initial spins are independent Bernoulli random variables with parameter $1/2$, and then evolve the system over $n^2/2$ units of time to achieve the steady state.
Next, in one set of simulations, we keep evolving the system, and make a histogram of $M_n(k)$ for $k=0,n,\dots,10^7n$, where $M_n(k)$ is the magnetization after $k$ units of time. This data are used to determine the distribution function of $M_n-\mu_0 n$.

In another set of simulations we obtain an estimate of $\left< (M_n(0)-M_n(t))^2\right>$ by averaging $(M_n(i)-M_n(i+j))^2$ for
$i=0,n+T,\dots, 10^3 (n+T)$ and $j=0,1,\dots,T$ in which $T=2n^{2/3}$ is the longest time difference over which we measure the correlation.

\begin{figure}
\begin{center}
\includegraphics{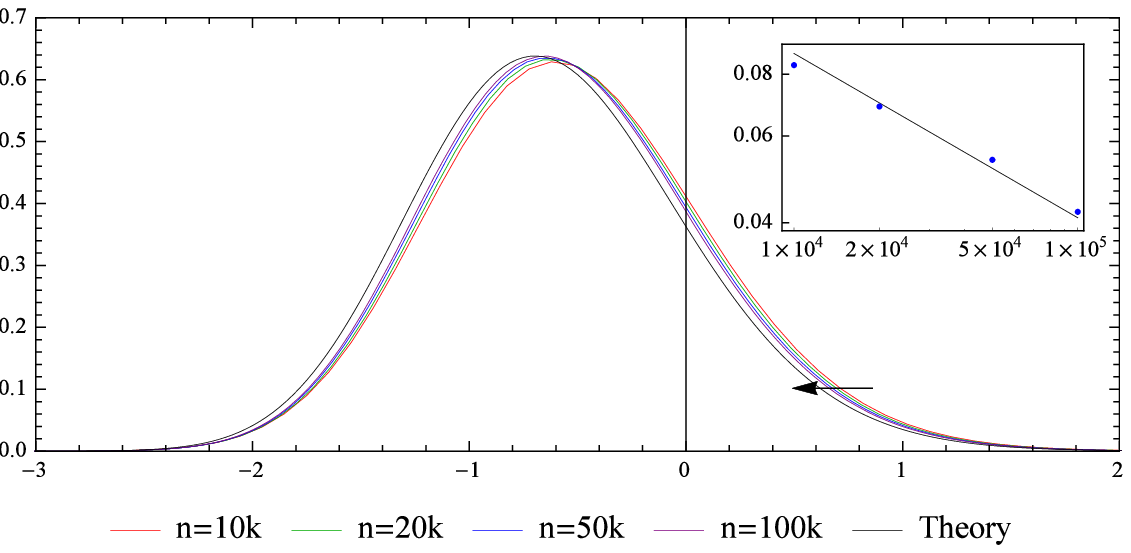}\quad
\includegraphics{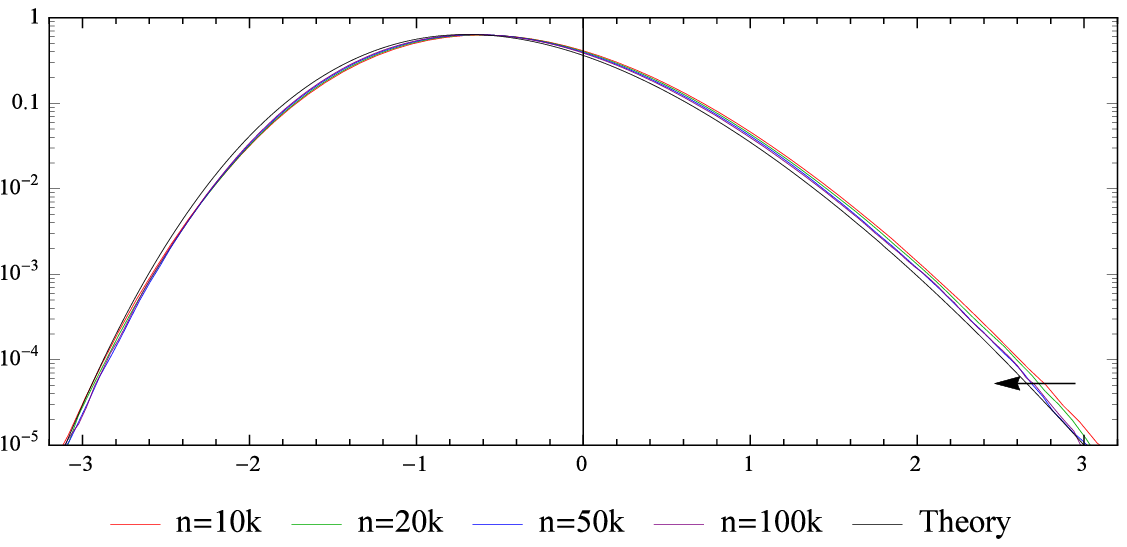}
\caption{Densities of $M_n^{\rm resc}$ for $n=10^4$, $2\times 10^4$, $5\times 10^4$, and $10^5$ compared with the theoretical prediction, that is, the density of $\frac12 \xi_{\rm GOE}$. The insert at the top figure is the log-log plot of the function $n\mapsto  \langle M_n^{\rm resc}\rangle- \tfrac12\langle \xi_{\rm GOE}\rangle$. The line has slope $-1/3$. The arrow indicates the shift of the curves as $n$ increases.}
\label{FigDensity}
\end{center}
\end{figure}

\begin{figure}
\begin{center}
\includegraphics{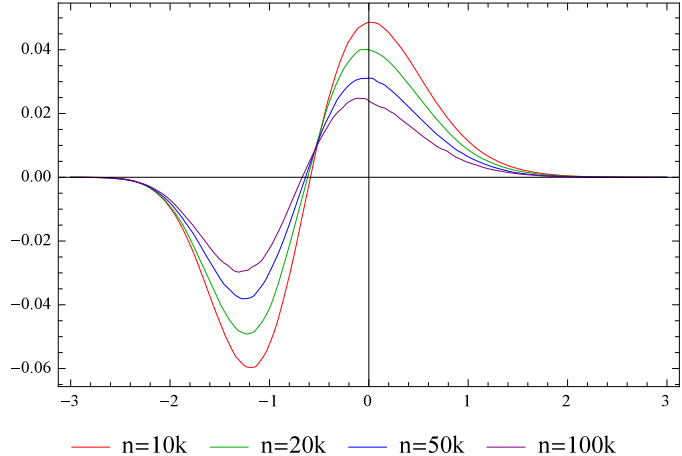}
\caption{Difference of the densities of $M_n^{\rm resc}$ for $n=10^4$, $2\times 10^4$, $5\times 10^4$, and $10^5$ and the theoretical prediction.}
\label{FigDensityDiff}
\end{center}
\end{figure}

We have made simulations for $\lambda=1/8$, a value at which the convergence with increasing system size is relatively fast, for $n=10^4$, $2\times 10^4$, $5\times 10^4$, and $10^5$.
We import the data sets in Mathematica and rescale them according to the theoretical predictions of (\ref{3.4}) and (\ref{3.11}). First we consider the scaling of the magnetization as
\begin{equation}
M_n^{\rm resc}=\frac{M_n-\mu_0 n}{(\tilde \Gamma n)^{1/3}}
\end{equation}
and compare its density with the one of $\frac12 \xi_{\rm GOE}$ (the data for $\xi_{\rm GOE}$ are taken from~\cite{PSKPZ}), see Figures~\ref{FigDensity} and~\ref{FigDensityDiff}. The agreement is remarkable and at first approximation one only sees a (non-random) shift of the distributions to the right, which goes to zero as $n^{-1/3}$ as observed previously in other models in the KPZ universality class, see~\cite{TS10,TS12,FF11,Tak12}.

Secondly, we focus at the covariance. Since our simulation is in steady state, we can derive the covariance from $\left< (M_n(0)-M_n(t))^2\right>$ simply by the relation
\begin{equation}
{\rm Cov}(M_n(0),M_n(t)) = {\rm Var}(M_n(0))-\tfrac12 \left< (M_n(0)-M_n(t))^2\right>.
\end{equation}
The value of ${\rm Var}(M_n(0))$ can be be obtained using the first set of data or by making an average over the region of times $t$ where $\left< (M_n(0)-M_n(t))^2\right>$ is constant. We used the latter approach, since it turns out to be less sensitive to long-lived correlations in the total magnetization, associated with the system's state close to the origin. The estimate of the variance has been made by averaging the values of $\tfrac12 \left< (M_n(0)-M_n(t))^2\right>$ for times $t\in [n^{2/3},2 n^{2/3}]$. In that region the theoretical prediction gives that the covariance (of the rescaled process) is about $10^{-6}$, which is much below the statistical noise that is about $10^{-3}$.

\begin{figure}
\begin{center}
\includegraphics{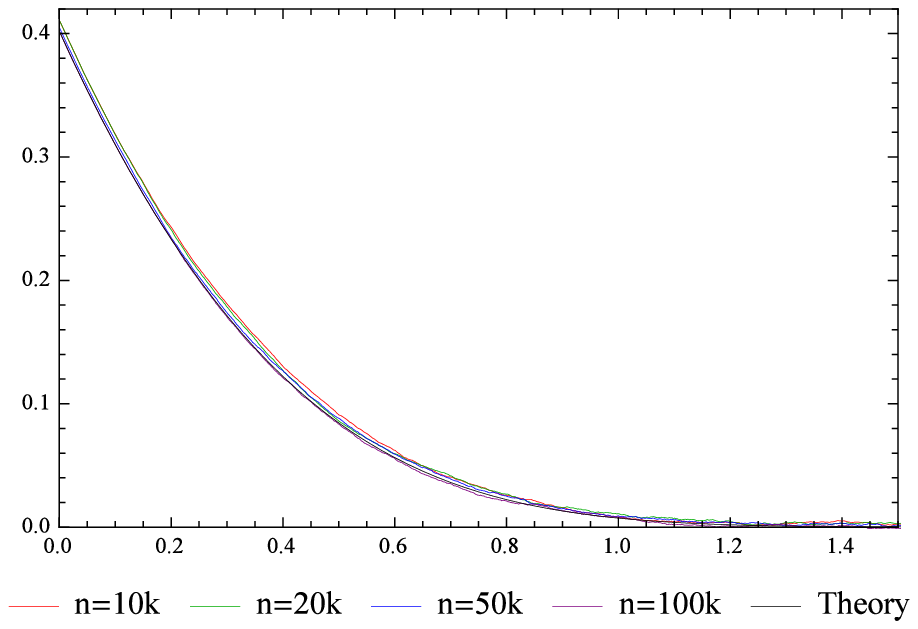}\quad
\includegraphics{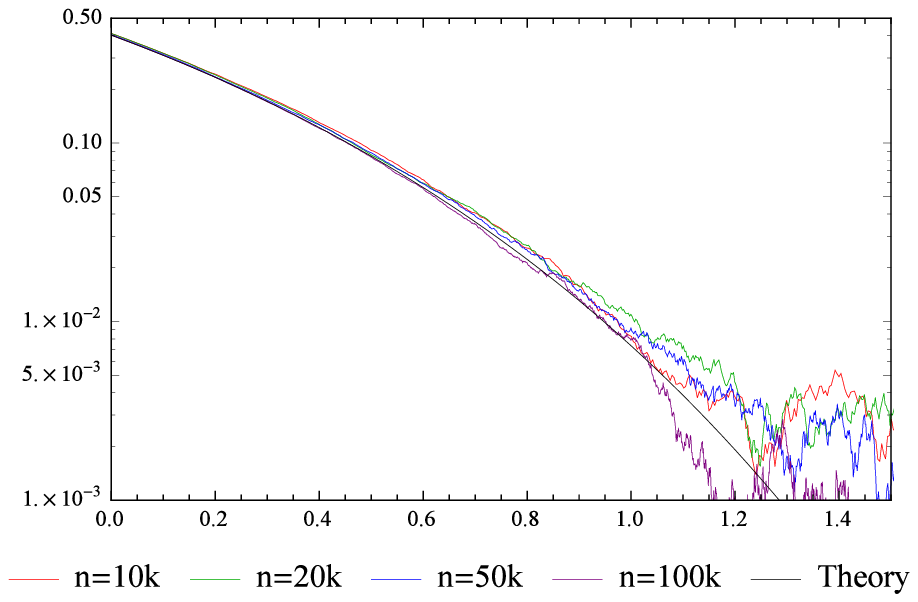}
\caption{Densities of ${\rm Cov}(M_n^{\rm resc}(0),M_n^{\rm resc}(t))$, $t\in [0,1.5]$, for $n=10^4$, $2\times 10^4$, $5\times 10^4$, and $10^5$ compared with the theoretical prediction $g_1(t)$.}
\label{FigCov}
\end{center}
\end{figure}

We considered the scaled process according to (\ref{3.11}), namely
\begin{equation}
M_n^{\rm resc}(t)=\frac{M_n(2 t (\tilde \Gamma n)^{2/3}/\tilde A)-\mu_0 n}{(\tilde \Gamma n)^{1/3}}.
\end{equation}
Using the approach described above, we determine the covariance of $M_n^{\rm resc}$ and plot it against the covariance $g_1(t)$ of the Airy$_1$ process, see Figure~\ref{FigCov}.

%%%%%%%%%%%%%%%%%%%%%%%%%%%%%%%%%%%%%%%%%%%%%%%%%%%%%%%%%%
\begin{table}
\begin{tabular}{|l|l|l|l|l|}
 \hline
 & Mean & Variance & Skewness & Kurtosis \\
 \hline
 $n=10\,000$ & $-0.5198; -14\%$ & $0.4335; 6.3\%$ & $0.2657; -9.4\%$ & $3.154; -0.33\%$ \\
 $n=20\,000$ & $-0.5344; -11\%$ & $0.4239; 3.9\%$ & $0.2757; -5.9\%$ & $3.159; -0.18\%$ \\
 $n=50\,000$ & $-0.5496; -9.0\%$ & $0.4162; 2.0\%$ & $0.2820; -3.8\%$ & $3.152; -0.40\%$ \\
 $n=100\,000$ & $-0.5612; -7.0\%$ & $0.4116; 0.9\%$ & $0.2897; -1.2\%$ & $3.168; \phantom{+}0.09\%$ \\
 $n=\infty$ & $-0.6033$ & $0.4080$ & $0.2931$ & $3.165$ \\
 \hline
\end{tabular}
\caption{Mean, Variance, Skewness, and Kurtosis of $M_n^{\rm resc}$ and their relative difference with the asymptotic values.}
\label{Table}
\end{table}

The precision in the agreement between theory and Monte Carlo data can be tested also through recording the higher order statistics, see Table~\ref{Table}. One expects that generically the $\ell$-th cumulant approaches its asymptotic value as $n^{-\ell/3}$. In particular the mean should have the slowest decay, consistent with our data.

\section{Conclusions}\label{sec5}
Using improved computer resources we have identified the distribution sampled in~\cite{BBLS96}, Fig. 3,
as the GOE Tracy-Widom edge distribution. In our figures there is no free scaling parameter. All model-dependent
parameters are computed from a sophisticated version of the KPZ scaling theory. One might wonder whether
similar properties hold for other 1D spin models with short range spin exchange dynamics. Such a model would have
a spin current $J(\mu)$ depending on the average magnetization $\mu$. We crucially used that $J(\mu) =0$
has a unique solution, $\mu_0$, with $|\mu_0| <1$. The case of multiple solutions has not been considered yet.
Furthermore we needed $J'(\mu_0) > 0$ corresponding to the right half lattice. If in addition $J''(\mu_0) \neq 0$,
the same properties as discussed in our note are predicted. If $J''(\mu_0) = 0$ but $J'''(\mu_0) \neq 0$, the variance grows as
$\sqrt{n}$ with logarithmic corrections. In principle also $J'''(\mu_0)$ could vanish. Then the asymptotics should behave
exactly as $\sqrt{n}$. The Toom spin model is singled out because it appears naturally from an underlying
cellular automaton.

\subsubsection*{Acknowledgements}
We thank Michael Pr\"{a}hofer, Jeremy Quastel, and Gene Speer for very helpful comments. The work was mostly done when the three last authors stayed at the Institute for Advanced Study, Princeton. We are grateful for the support. The work of JLL was supported by NSF Grant DMR-1104501. PLF was supported by the German Research Foundation via the SFB 1060--B04 project.

\end{document}